# Extending Reflectometry Range: A Zero-Crossing Algorithm for Thick Film Metrology

**Zimu Zhou\*†, Enrique A. Lopez-Guerra†, Iulica Zana, Vu Nguyen, Nguyen Quoc Huy Tran, Bojun Zhou, Gary Qian, Michael Kwan, Peter Wilkens, Chester Chien**

*Western Digital Corp. (USA)*
*Hardware Development, 5601 Great Oaks Pkwy, San Jose, CA, 95119*
†*The authors contributed equally to this work*
\* *zimu.zhou@wdc.com*

**Abstract:** Accurate and high-efficiency film metrology remains a key challenge in High-Volume Manufacturing (HVM), where conventional spectroscopic reflectometry and white light interferometry (WLI) are either limited by model dependence or throughput. In this work, we extend the measurable film-thickness range of reflectometry to at least 50 μm through a new model-free algorithm, the Linearized Reflectance Zero-Crossing (LRZ) method. The approach builds upon the previously reported Linearized Reflectance Extrema (LRE) technique but eliminates the sensitivity to spectral sampling and fringe attenuation that degrade performance in the thick-film regime. By linearizing phase response and extracting zero-crossing positions in wavenumber space, LRZ provides robust and repeatable thickness estimation without iterative fitting, achieving comparable accuracy with much higher computational efficiency than conventional model-based methods. Validation using more than 80 measurements on alumina films over NiFe substrates shows excellent correlation with WLI (r = 0.97) and low gauge repeatability and reproducibility (GR&R < 3%). Moreover, LRZ achieves an average Move-Acquire-Measure (MAM) time of approximately 2 s, which is about 7 times faster than WLI. The proposed method enables fast, accurate, and model-independent optical metrology for thick films, offering a practical solution for advanced HVM process control.

## 1. Introduction

As microelectronic devices continue to shrink in size and increase in complexity, the demand for precise and reliable thin film metrology in modern High-Volume Manufacturing (HVM) has grown substantially. Accurate thickness control of functional layers — such as dielectrics[1], polymers, nanostructured template[2], and protective coatings — is critical for maintaining device performance, yield, and long-term reliability.[3] Among various metrology techniques, optical reflectometry has gained wide adoption in production environments due to its non-destructive nature, high throughput, and ease of integration with semiconductor process tools.[4] Reflectometry is an optical metrology technique that measures thin film thickness by analyzing the interference patterns generated by light reflected from the sample surface and underlying layers. The resulting spectral data is fitted to theoretical models, providing accurate and non-destructive calculation of film thickness. In our previous work[5], we proposed a novel model-free methodology, the Linearized Reflectance Extrema (LRE) approach, which directly calculates film thickness from the reflectance extrema, eliminating iterative processes and significantly improving both computational throughput and measurement repeatability in high-volume manufacturing environments.

However, both iterative and LRE approaches will eventually encounter challenges as film thickness increases beyond certain limits. Traditional iterative reflectometry methods are generally limited to film thicknesses below approximately 20–30 µm. This limitation arises from difficulties in resolving high-order spectral fringes, increased ambiguity in layer modeling, and decreased sensitivity at larger optical path differences. Similarly, the LRE method also faces restrictions for very thick films. As the film gets thicker, the fringe frequency in the reflectance spectrum increases, while the sampling rate remains fixed by the spectrometer's resolution. This leads to unreliable extraction of reflectance extrema, introducing phase errors and ultimately causing significant inaccuracies in thickness calculation. Fourier-based methods like the Discrete Fourier Transform (DFT) have been used for thick film reflectometry, but they come with several practical limitations specific to this application. As Quinten points out[6], the Fast Fourier Transform (DFT's most widely used algorithm) assumes equidistant sampling, yet reflectance spectra are typically sampled uniformly in wavelength ($\lambda$), not in wavenumber ($k$), which is the physically relevant variable. This mismatch requires interpolation, introducing further inaccuracies. To address this, Quinten proposes combining Fast Fourier Transform (FFT) with grid search and nonlinear regression, though this adds computational complexity. Additionally, the discrete and finite nature of the signal leads to well-known caveats in Fourier analysis: aliasing and spectral leakage.[7–9] These constraints restrict the application of reflectometry for monitoring thick transparent or semi-transparent films, which are becoming increasingly important in advanced manufacturing processes, such as wafer bonding[10], 3D packaging[11], and flexible electronics[12]. We will demonstrate some non-ideal measurement examples later to highlight these challenges in section 3.1.

Nowadays, many metrology applications in Hard-Disk Drive (HDD) fabrication require measuring the thickness of uniform alumina films over NiFe substrates, with thicknesses ranging from a few hundred nanometers up to 60 µm. Accurately measuring such a broad range has been challenging due to hardware and algorithmic limitations. In this work, we present a novel algorithmic enhancement to standard reflectometry that extends the measurable film thickness range to at least 50 µm—without requiring any hardware modification. Our approach incorporates improved spectral modeling, fringe analysis, and curve-fitting techniques to extract reliable thickness values from high-frequency spectral patterns. The method is validated on thick transparent films using gauge repeatability and reproducibility (GR&R) testing across more than 300 measurements, demonstrating high consistency and robustness.

This purely software-based solution significantly expands the application range of reflectometry for HVM. It offers a cost-effective, inline-compatible alternative to more complex or slower systems such as White Light Interferometry (WLI), confocal microscopy, or cross-sectional FIB/SEM.

## 2. Materials and methods

*2.1 Reflectometry*

Wafer-level reflectometry measurements were performed using a commercial 200 mm system (NANOSPEC® 9100 Series, P/N 7500-1214-C). The instrument employs both a visible (VIS) halogen lamp and an ultraviolet (UV) deuterium lamp, covering wavelength ranges of 210 – 750 nm and 400 – 780 nm, respectively. The detector operates over approximately 190–780 nm. Measurements are acquired at normal incidence, with a nominal spectral bandwidth of ~4 nm and a wavelength sampling interval of ~0.58 nm.

## 2.2 White Light Interferometry

White light interferometry measurements are performed using a Bruker Contour GT-X optical profiler operating in Vertical Scanning Interferometry (VSI) mode. The system uses a broadband LED light source to illuminate the sample. Film thickness measurements are conducted using the thick film algorithm, suitable for semi-transparent films with thicknesses greater than 2 µm. The algorithm identifies modulation envelopes corresponding to the film surface and the substrate interface. When both envelopes exceed the VSI modulation threshold, their separation is computed and divided by the refractive index to determine the film thickness. No topography is measured in this mode. The system includes motorized X/Y/Z, and tip/tilt stages for automated site acquisition.

## 2.3 Sampling plan

To evaluate the accuracy of the reflectometry algorithm, measurements were performed on a patterned 200 mm wafer using a set of available test devices. A total of 9 sites were selected, distributed across the wafer to include the center, intermediate radius, and near-edge regions. The selection provides coverage along both horizontal and vertical axes, as well as diagonal directions, allowing sensitivity to radial and azimuthal variations in film thickness. At each site, both tools performed nine repeated measurements, resulting in a total of 81 data points per tool. This approach ensures statistical robustness while maintaining direct site-to-site comparability between the two metrology techniques.

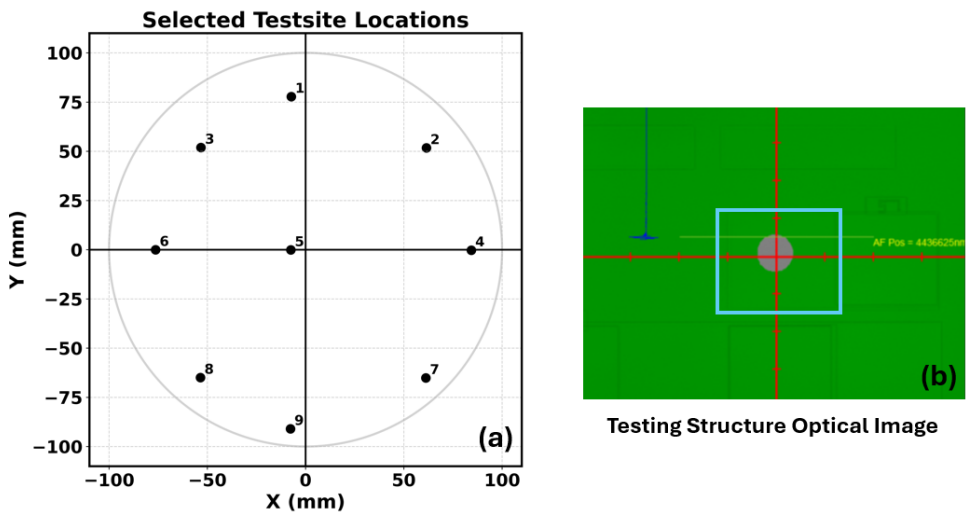

Figure 1a. Sampling map on the 200 mm wafer. 1b. Optical image of the testing structure.

*2.4. GR&R*

Repeatability (EV) represents the variation observed when repeated measurements are performed on the same part using a single tool under identical conditions. Its mathematical definition is provided in Eq. (1) and (2)

$$EV = 6 \cdot \sqrt{VAR(within)} \tag{1}$$

$$VAR(within) = \frac{1}{N} \sum_{i=1}^{N_{PART}} \sum_{j=1}^{N_{OPERATOR}} \sigma_{ij}^2 \tag{2}$$

GR&R (Gage Repeatability and Reproducibility) refers to the actual variation in measurement caused by the measurement system. Since the present study involves only one measurement tool and a fixed operator, the appraiser variation (AV) is negligible, and thus the gauge R&R (GR&R) effectively reduces to EV.

Part Variation (PV) PV quantifies the variation in measurement results that arises from actual differences among the test samples or sites. In other words, it reflects the true physical variability of the film thickness across the wafer, independent of the measurement system.

$$PV = \sqrt{\frac{1}{N_{PART} - 1} \sum_{i=1}^{N_{PART}} (\bar{X}_i - \bar{X})^2} \tag{3}$$

Total Variation (TV) TV represents the quadratic sum of part-to-part variation and measurement system variation.

$$TV = \sqrt{EV^2 + PV^2} \tag{4}$$

GR&R% quantifies how much of the total variation in a measurement system comes from the gauge itself,

$$GR\&R\% = \frac{GR\&R}{TV} \tag{5}$$

Per the Automotive Industry Action Group (AIAG) Measurement Systems Analysis (MSA) guideline, a GR&R% below 10% denotes an acceptable measurement system, while values between 10% and 30% may be acceptable depending on the specific application.[13,14]

## 3. Results

*3.1 Limitations of LRE Method and Iterative Fitting Method*

In our previous work, the Linearized Reflectance Extrema (LRE) method demonstrated reliable performance for films up to approximately 18 μm. However, this approach becomes less suitable for thicker films, primarily due to the spectrometer's discrete sampling rate and optical attenuation at large optical paths. As the film thickness increases, the interference fringes become densely packed and attenuated, making it difficult to precisely identify the true peak and valley positions. This leads to phase ambiguity and cumulative phase-shift errors, which significantly degrade the accuracy and repeatability of thickness extraction.

Figures 2a and 2b present a GR&R analysis of the LRE method applied to a ~54 μm film. Figure 2a shows the measured thickness variability across 9 wafer sites, while the lower panel (2b) displays the corresponding within-site standard deviation. Despite 9 repeated measurements per site, the LRE method exhibits substantial site-to-site fluctuations and large within-site variation, yielding a GR&R% of 75.9%, indicating that the method cannot effectively resolve part-to-part differences. These results confirm that the extrema-based algorithm becomes unreliable for thick-film applications, where dense oscillations and optical attenuation hinder precise extrema detection.

To provide a baseline comparison, we apply the standard model-based spectroscopic reflectometry method, where each measured spectrum is fitted against a theoretical target spectrum derived from a predefined optical model. The material stack is modeled with dispersion relations for the complex refractive index $N(\lambda)+iK(\lambda)$: the bottom NiFe substrate is described using a Lorentz oscillator model, and the film layer is modeled using a Cauchy dispersion relation, appropriate for transparent materials in the visible range. The fitting procedure minimizes the squared error between the measured and modeled spectra by floating the film thickness. To try to mitigate the issue of multiple local minima[5], we use a refined search grid ranging from 40 μm to 60 μm with 600 grid points.

Figure 2c and 2d show the corresponding GR&R analysis for the conventional iterative method. Compared to the LRE results, the iterative approach demonstrates significantly improved repeatability and stability, with a GR&R of only 5.8%, indicating a highly capable measurement system. The within-site standard deviations are typically below 120 nm, and most sites exhibit consistent fitting convergence. However, despite the improved measurement precision, this approach is computationally expensive—each site requires approximately 60 ms of algorithm processing time, which is 20–30× slower than the non-iterative methods such as LRE or LRZ. Consequently, while the optical model fitting yields acceptable measurement performance, its high computational overhead limits practical throughput in production environments.

Overall, these findings highlight the trade-offs between the LRE and conventional iterative methods for thick-film reflectometry. The LRE method suffers from sampling-induced phase errors and attenuation-driven instability, while the iterative fitting approach achieves better precision at the expense of speed, robustness, and model dependency. These limitations collectively motivate the development of the Linearized Reflectance Zero-crossing (LRZ) technique introduced in this work, which combines the phase-tracking robustness of a model-free algorithm with the precision of a phase-linearized framework, enabling accurate and repeatable measurement across a wide thickness range.

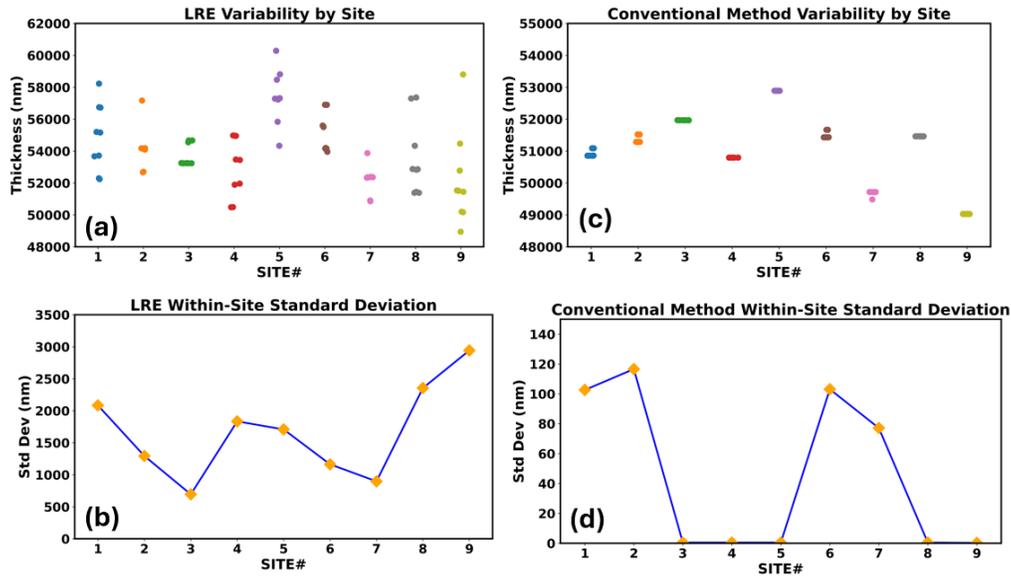

Figure 2a and 2b. LRE variability and within-site standard deviation plots. 2c and 2d. Model-based variability and within-site standard deviation plots.

### 3.2 Proposed LRZ Mathematical Framework

The challenges discussed in the previous section clearly underscore the limitations of both the iterative model-based and LRE algorithms in handling thick-film measurements (>~35 μm). The iterative approach, while accurate, is computationally intensive and model-dependent, whereas the LRE method suffers from phase instability and poor repeatability under dense fringe conditions. To overcome these challenges, we developed a new algorithm derived from the LRE framework, designed to preserve its model-free nature while improving phase linearity and measurement robustness for thick films. This approach, termed the Linearized Reflectance Zero-Crossing (LRZ) method, is described in detail in the following section.

Reflectometry relies on analyzing the interaction of light with thin films, using the reflected light spectrum to infer material properties like thickness and refractive index. It is well known that the spectral reflectance can be calculated using the Airy equations Eq. (6)

$$R = |r_{total}|^2 = r \times r^* = \left| \frac{r_{01} + r_{12} \exp(-i2\phi)}{1 + r_{01} r_{12} \exp(-i2\phi)} \right|^2 \qquad (6)$$

Eq. (7) represents the phase shift

$$\phi(k) = kd N_1(k) \qquad (7)$$

Where

$k$: wavenumber

$$k = \frac{2\pi}{\lambda}$$

$d$: film thickness.

$N_1(k)$: effective refractive index of the thin film

In many semiconductor and HDD applications, thin film materials such as aluminum oxides, silicon oxides and photoresist have very low extinction coefficients[15]. Therefore, it is safe to assume that the refractive index of the measured thin film $N_1$, is a real number. Subsequently, $R$ can be expressed in terms of Eq (8) after expanding both numerator and denominator:

$$R = \frac{r_{01}^2 + |r_{12}|^2 + 2r_{01}(r_{12R}\cos(2\phi) + r_{12I}\sin(2\phi))}{1 + r_{01}^2|r_{12}|^2 + 2r_{01}(r_{12R}\cos(2\phi) + r_{12I}\sin(2\phi))} \tag{8}$$

Where $r_{12R}$ and $r_{12I}$ refer to the real and imaginary Fresnel reflectance coefficients, respectively. Or, in a simplified expression that will become handy for subsequent derivations in Eq (9):

$$R = \frac{A + B\cos(\psi)}{C + B\cos(\psi)} \tag{9}$$

Where

$$\psi = 2\,k\,N_1 d + \delta$$

The mean reflectance $\bar{R}$ over one full oscillation period can be expressed in closed form using the parameters A, B, C, which depend only weakly on the phase term $\psi$.

$$\bar{R} = \frac{1}{2\pi}\int_0^{2\pi,} R(\psi)\,d\psi = \frac{A\,-\,C\,+\,\sqrt{\{C^2 - B^2\}}}{\sqrt{\{C^2 - B^2\}}} \tag{10}$$

As a result, $\bar{R}$ is effectively a constant and does not vary significantly with $\psi$. This indicates that dispersion has little influence on the overall shape of the reflectance, especially for thicker films. Subtracting this constant from $R(\psi)$ yields a mean-centered waveform, where the zero-crossings carry the information about film thickness.

The zero-crossing condition is defined by

$$R(\psi_0) - \bar{R} = 0 \tag{11}$$

Which can be rearranged as

$$\bar{R} = \frac{A + B\cos(\psi_0)}{C + B\cos(\psi_0)} \tag{12}$$

Clearing the denominator leads to

$$A - \bar{R}C = (\bar{R} - 1)B\cos(\psi_0) \tag{13}$$

Defining

$$e = \frac{A - \bar{R}C}{B(\bar{R} - 1)}, |e| \le 1 \tag{14}$$

The condition simplifies to

$$\cos(\psi_0) = e \qquad (15)$$

All solutions are given by

$$\psi_0^{(n)} = \arccos(e) + \pi n, n \in \mathbb{Z} \qquad (16)$$

Substituting into the phase expression, leads to the zero-crossing wavenumber as

$$k_n = \frac{\arccos(e) - \delta}{2N_1 d} + \frac{\pi}{2N_1 d} n \qquad (17)$$

This shows a linear dependence of $k_n$ on the index $n$. Differentiating

$$\frac{d(k_n)}{dn} = \frac{\pi}{2N_1 d} \qquad (18)$$

so that the optical path length follows as

$$N_1 d = \frac{\pi}{2 \cdot \frac{d(k_n)}{dn}} \qquad (19)$$

Thus, the slope of a straight-line fit of $k_n$ versus $n$ directly yields the optical thickness, from which the physical film thickness $d$ can be determined. This assumption is well justified for aluminum oxide and many other semi-transparent dielectric films. Under this condition, $N_1$ may be treated as a known material constant. Consequently, the film thickness $d$ can be determined directly from the slope $\frac{d(k_n)}{dn}$ and the specified value of $N_1$.

Figure 3(a) showed the detrended reflectance spectrum within the wavenumber range of 0.0097–0.0100 nm⁻¹. The oscillatory interference fringes are preserved, and the interpolation enables accurate determination of the zero-crossing locations (red crosses) between discrete sampling points (orange circles). These zero crossings correspond to the condition $\Delta\psi = n\pi$, which is directly proportional to the film thickness. Figure 3(b) showed the extracted wavenumber positions of the first 20 zero crossings as a function of zero-crossing index. The linear regression yields an excellent coefficient of determination ($R^2 = 0.999$), confirming the linear relationship between fringe order and wavenumber. The slope $\Delta k$ provides a direct measure of the optical path length within the film, thereby enabling thickness extraction with high accuracy. The near-perfect linearity demonstrates that dispersion effects are negligible within this spectral window, validating the robustness of the zero-crossing method. More importantly, this result confirms that the proposed zero-crossing approach can be reliably applied to extend reflectometry to the tens-of-microns thickness regime, while providing consistency for cross-validation against WLI measurements.

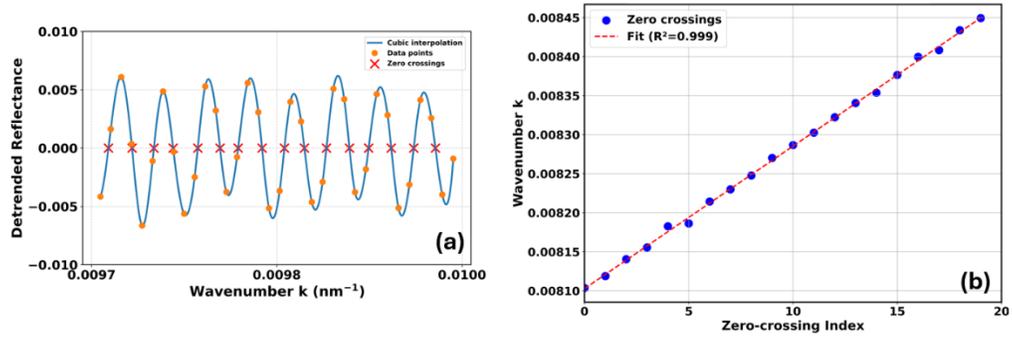

Figure 3a. Illustration of proposed LRZ algorithm zero-point extraction (red cross). Figure 3b Linear regression of zero-crossing order versus wavenumber, demonstrating phase linearity ($R^2 = 0.999$).

## 4. Results and Discussion

In the experiment, a total of 81 test sites were measured using both reflectometry and white light interferometry (WLI). The measurements were collected at identical test locations, evenly distributed across the wafer, to evaluate the thickness sensitivity and measurement consistency across different wafer positions.

Figure 4 shows the correlation between film thickness measured by reflectometry and white light interferometry (WLI) at 81 test sites across the wafer. An orthogonal regression yields a correlation coefficient of $r = 0.9689$, indicating excellent agreement between the two metrology techniques. The close alignment of data points along the unity line (red dashed) demonstrates that the zero-crossing method in reflectometry provides consistent thickness values when benchmarked against WLI, which is the established standard for this thickness regime.

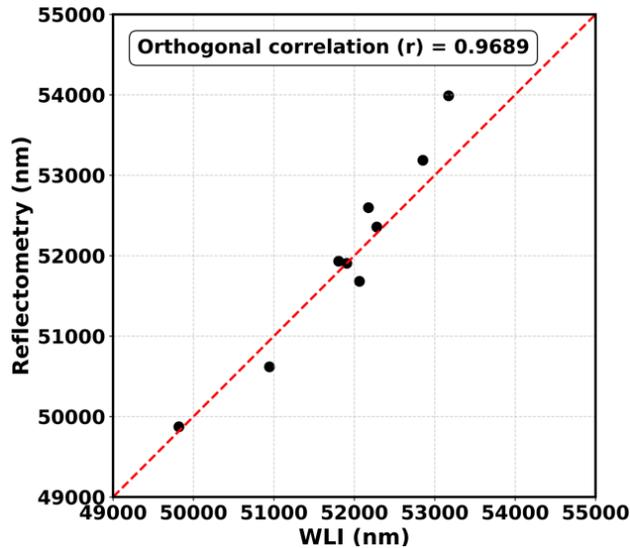

Figure 4 Correlation between WLI and reflectometry measurement using LRZ algorithm.

This strong correlation validates the robustness of the proposed reflectometry approach for films in the tens-of-microns range. Small deviations from the unity line are expected and generally acceptable for film thickness measurements in the ~50 μm range, as the observed differences are within ~1%. This level of variation is well within the tolerance of both reflectometry and WLI, and is more than sufficient for real-world high-volume manufacturing metrology applications. Moreover, unlike our previous work where reliable reflectometry data up to 18 μm thickness could be demonstrated but without an independent metrology reference, the present study incorporates direct cross-validation with WLI. This confirms that reflectometry can serve as a reliable, non-contact alternative for thick-film measurements and provides a viable path for extending the application range of optical reflectometry in semiconductor metrology.

Figure 5 is the comparison of site-to-site variability between WLI and the proposed LRZ method. Figures (a) and (c) show the measured film thickness values at each wafer site, where both techniques exhibit consistent site-dependent variations. Figures (b) and (d) present the corresponding within-site standard deviations for WLI and LRZ, respectively.

Both methods demonstrate good repeatability, with typical within-site standard deviations below 50 nm (<0.1% of the ~50 μm film thickness). Although LRZ exhibits slightly larger scatter at certain sites (e.g., site 6), the overall variability remains within acceptable limits for thick-film metrology. These results indicate that the reflectometry-based LRZ approach achieves comparable precision to WLI, confirming that the method is not only accurate on average, but also robust at the site-to-site level. Such repeatability is critical for ensuring metrology reliability in high-volume manufacturing.

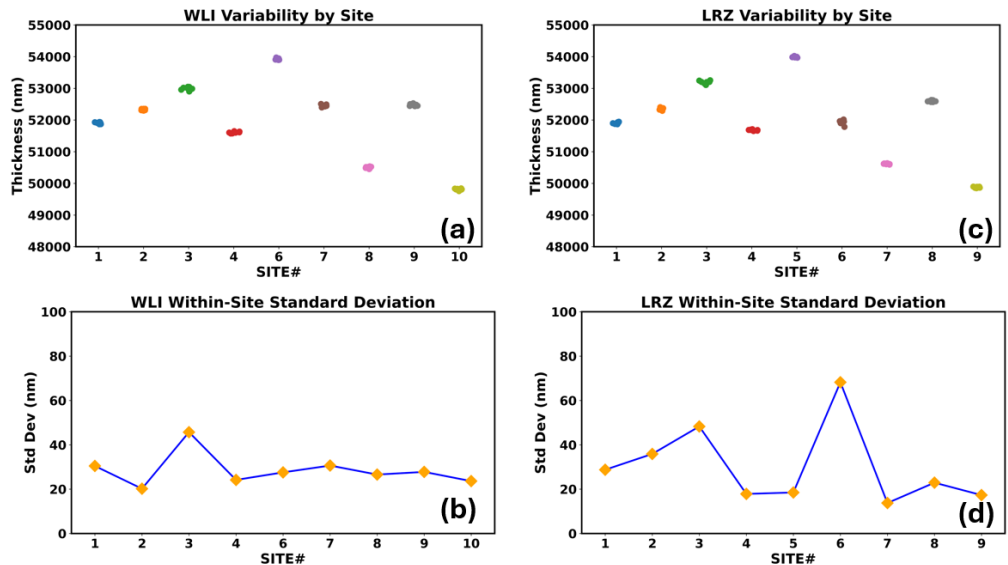

Figure 5a and 5b. WLI variability and within-site standard deviation plots. 5c and 5d. LRZ variability and within-site standard deviation plots.

Table 1 shows the comparison of gauge repeatability and reproducibility (GR&R) metrics between the LRZ method and WLI. Both methods show nearly identical total variation (1244.7 nm vs 1241.2 nm) and low GR&R percentages (2.78% for LRZ and 2.37% for WLI), confirming that LRZ achieves comparable measurement accuracy to the established WLI reference. The repeatability component (EV) is slightly higher for LRZ (34.6 nm vs 29.4 nm), but the difference remains negligible relative to the ~50 μm film thickness. Overall, the total variation is ~1245 nm, of which less than 3% originates from GR&R. This indicates that the majority of the observed variation arises from true wafer nonuniformity (PV), while the measurement system error (EV/ GR&R) remains negligible in comparison

More importantly, LRZ demonstrates a significant throughput advantage, with an average per-site processing time of approximately 2 s, compared to ~15 s for WLI. This represents a ~7× speed improvement while maintaining equivalent accuracy. Considering that typical HVM metrology involves > 50 sampling sites per wafer, the time savings quickly accumulate — reducing total measurement time per wafer from more than 12 minutes to under 2 minutes. This substantial efficiency gain highlights the potential of LRZ reflectometry as a practical and scalable alternative to WLI for high-volume manufacturing applications.

Table 1 GR&R Results and Throughput Comparison

| Test Item | LRZ | WLI |
|---|---|---|
| Repeatability (EV) | 34.6 nm | 29.4 nm |
| Total Variation (TV) | 1244.7 nm | 1241.2 nm |
| GR&R % | 2.78% | 2.37% |
| Avg MAM (Move-Acquire-Measure) Time | ~2s | ~15s |

Another key advantage of the LRZ method over traditional model-based iterative approaches is its resilience to process excursions. In iterative reflectometry, the solution space is constrained by the predefined grid range. For example, if the search space is limited to 40–60 μm, any measurement outside this range—such as a process excursion resulting in a 30 μm or 70 μm film—will fail to converge, leading to invalid or missing data. In contrast, LRZ does not rely on bounded parameter grids. As long as the reflectance spectrum contains a sufficient number of zero crossings, the algorithm can extract the slope and determine the film thickness. This makes LRZ inherently more robust in detecting unexpected thickness variations, which is critical for monitoring excursions in high-volume manufacturing environments

## 5. Conclusion

The Linearized Reflectance Zero-Crossing (LRZ) algorithm successfully extends the measurable thickness range of optical reflectometry to at least 50 μm without requiring any hardware modification. By linearizing the phase information and tracking zero-crossing positions in wavenumber space, LRZ provides accurate and repeatable film-thickness estimation while maintaining model-free operation. Experimental validation against white light interferometry (WLI) across 81 wafer sites demonstrates excellent correlation (r = 0.97) and low GR&R (< 3%), confirming measurement equivalence. Moreover, LRZ achieves roughly a seven-fold throughput improvement (~2s per site) compared to WLI, making it a highly efficient and scalable metrology solution for high-volume manufacturing. This purely software-based enhancement offers a practical path to integrate fast, accurate, and model-independent thick-film metrology into production environments.

Future work may focus on a detailed comparison between the proposed LRZ and the previously developed Linearized Reflectance Envelope (LRE) method to further clarify their respective performance domains and limitations. Additionally, systematic exploration of the upper measurable thickness limit of LRZ, potentially beyond 60 µm, will help establish its full operational range and identify conditions where fringe density or detector resolution may become limiting factors. Extending LRZ to multilayer or dispersive films could also broaden its applicability in advanced process control.

**Captions:**

Figure 1a. Sampling map on the 200 mm wafer. 1b. Optical image of the testing structure.

Figure 2a and 2b. LRE variability and within-site standard deviation plots. 2c and 2d. Model-based variability and within-site standard deviation plots.

Figure 3a. Illustration of proposed LRZ algorithm zero-point extraction (red cross). Figure 3b Linear regression of zero-crossing order versus wavenumber, demonstrating phase linearity ($R^2$ = 0.999).

Figure 4 Correlation between WLI and reflectometry measurement using LRZ algorithm.

Figure 5a and 5b. WLI variability and within-site standard deviation plots. 5c and 5d. LRZ variability and within-site standard deviation plots.

**Funding:** This research did not receive any funding from external sources outside of Western Digital Corporation

**Acknowledgments:** The authors would like to thank Yingjian Chen, Ian McFadyen and Michael Parker for their guidance and expertise.

**Conflicts of Interest:** The authors declare no conflict of interest.

**Data Availability:** The reflectometry spectra and WLI data analyzed in this study are not publicly available due to commercial confidentiality. However, the data may be shared upon reasonable request for scientific research purposes, subject to approval by Western Digital Corporation.

**Reference**

1.  J. De Roo et al., "Synthesis of Phosphonic Acid Ligands for Nanocrystal Surface Functionalization and Solution Processed Memristors," Chemistry of Materials **30**(21), 8034–8039, American Chemical Society (2018) [doi:10.1021/acs.chemmater.8b03768].

2.  Z. Zhou and S. S. Nonnenmann, "Progress in Nanoporous Templates: Beyond Anodic Aluminum Oxide and Towards Functional Complex Materials," Materials **12**(16), 2535 (2019) [doi:10.3390/ma12162535].

3.  "International Technology Roadmap for Semiconductors 2.0" (2015).

4.  O. Stenzel and M. Ohlídal, "Optical Characterization of Thin Solid Films."

5.  Z. Zhou et al., "Model-free methodology for thin film thickness analysis by spectroscopic reflectometry," Journal of Micro/Nanopatterning, Materials, and Metrology **24**(02) (2025) [doi:10.1117/1.JMM.24.2.024001].

6.  M. Quinten, "On the use of fast Fourier transform for optical layer thickness determination," SN Appl Sci **1**(8), Springer Nature (2019) [doi:10.1007/s42452-019-0866-9].

7.  E. A. López-Guerra, B. Uluutku, and S. D. Solares, "Insights about aliasing and spectral leakage when analyzing discrete-time finite viscoelastic functions," Reports in Mechanical Engineering **4**(1), 104–120, Regional Association for Security and crisis management (2023) [doi:10.31181/rme040129072023lg].

8.  R. G. Lyons, *Understanding Digital Signal Processing Third Edition*, Pearson (2010).

9.  Alan V. Oppenheim and Ronald W. Schafer, *Discrete-Time Signal Processing*, 1st ed., Marcia Horton, Ed., Upper Saddle River (1989).

10. N. G. Orji et al., "Metrology for the next generation of semiconductor devices," Nature Electronics **1**(10), pp. 532–547, Nature Publishing Group (2018) [doi:10.1038/s41928-018-0150-9].

11. Y.-S. Ku et al., "Machine learning-based 3D TSV metrology for advanced packaging," 21 February 2025, 137, SPIE-Intl Soc Optical Eng [doi:10.1117/12.3051271].

12. Z. Zhou et al., "Memristive Behavior of Mixed Oxide Nanocrystal Assemblies," ACS Appl Mater Interfaces **13**(18), 21635–21644 (2021) [doi:10.1021/acsami.1c03722].

13. A. A. Deshpande et al., "Applications of gage reproducibility & repeatability (GRR): Understanding and quantifying the effect of variations from different sources on a robust process development," Org Process Res Dev **18**(12), 1614–1621, American Chemical Society (2014) [doi:10.1021/op5002935].

14. Automotive Industry Action Group, *Measurement Systems Analysis (MSA) Reference Manual*, 4th ed., AIAG, Southfield, Michigan (2010).

15.     C. Kaiser et al., "Determining Ultralow Absorption Coefficients of Organic Semiconductors from the Sub-Bandgap Photovoltaic External Quantum Efficiency," Adv Opt Mater **8**(1), Wiley-VCH Verlag (2020) [doi:10.1002/adom.201901542].